\documentclass[aps, pra, 10pt, superscriptaddress, twocolumn,floatfix]{revtex4-2}
\usepackage{bbold}
\usepackage{graphicx}
\usepackage{amsmath,amssymb,amsfonts}
\usepackage[percent]{overpic}
\usepackage{braket}
\usepackage{booktabs}
\usepackage{bm}
\usepackage[breaklinks=true,colorlinks,citecolor=blue,linkcolor=blue,urlcolor=blue]{hyperref}
\usepackage[usenames, dvipsnames]{color}
\usepackage{mathrsfs}
\usepackage[normalem]{ulem}

\begin{document}
\title{Reinforcement learning optimization of the charging of a Dicke quantum battery}
\author{Paolo Andrea Erdman}
\altaffiliation{These two authors contributed equally.\\
\href{mailto:p.erdman@fu-berlin.de}{p.erdman@fu-berlin.de}\\
\href{mailto:gian-marcello.andolina@college-de-france.fr}{gian-marcello.andolina@college-de-france.fr}}
\affiliation{Freie Universit{\" a}t Berlin, Department of Mathematics and Computer Science, Arnimallee 6, 14195 Berlin, Germany}
\author{Gian Marcello Andolina}
\altaffiliation{These two authors contributed equally.\\
\href{mailto:p.erdman@fu-berlin.de}{p.erdman@fu-berlin.de}\\
\href{mailto:gian-marcello.andolina@college-de-france.fr}{gian-marcello.andolina@college-de-france.fr}}
\affiliation{ICFO-Institut de Ci\`{e}ncies Fot\`{o}niques, The Barcelona Institute of Science and Technology, Av. Carl Friedrich Gauss 3, 08860 Castelldefels (Barcelona),~Spain}
\affiliation{JEIP, UAR 3573 CNRS, Coll\`ege de France, PSL Research University, F-75321 Paris, France}

\author{Vittorio Giovannetti}
\affiliation{NEST, Scuola Normale Superiore and Istituto Nanoscienze-CNR, I-56126 Pisa,~Italy}
\author{Frank No{\'e}}
\email{frank.noe@fu-berlin.de}
\affiliation{Microsoft Research AI4Science, Karl-Liebknecht Str. 32, 10178 Berlin, Germany}
\affiliation{Freie Universit{\" a}t Berlin, Department of Mathematics and Computer Science, Arnimallee 6, 14195 Berlin, Germany}
\affiliation{Freie Universit{\" a}t Berlin, Department of Physics, Arnimallee 6, 14195 Berlin, Germany}
\affiliation{Rice University, Department of Chemistry, Houston, TX 77005, USA}

\begin{abstract}
Quantum batteries are energy-storing devices, governed by quantum mechanics, that promise high charging performance thanks to collective effects. Due to its experimental feasibility, the Dicke battery - which comprises $N$ two-level systems coupled to a common photon mode - is one of the most promising designs for quantum batteries. However, the chaotic nature of the model severely hinders the extractable energy (ergotropy). Here, we use reinforcement learning to optimize the charging process of a Dicke battery either by modulating the coupling strength, or the system-cavity detuning.
We find that the ergotropy and quantum mechanical energy fluctuations (charging precision) can be greatly improved with respect to standard charging strategies by countering the detrimental effect of quantum chaos. 
Notably, the collective speedup of the charging time can be preserved even when nearly fully charging the battery.
\end{abstract}

\maketitle

{\it Introduction.---}
It is believed that eventually quantum effects, such as entanglement and coherence, could be used to perform certain tasks that cannot be performed by a classical machine. Theoretical examples of that are known, for example, in the fields of computation \cite{nielsen2011} or cryptography \cite{gisin2002}.
Thermodynamics is an empirical theory, developed in the 19th century, that studies the transformation of energy into heat and work \cite{Fermi_Book}. 
Given the role that thermodynamics played in the industrial revolution, it is natural to ask whether quantum resources can be exploited to improve thermodynamic performances \cite{Pekola,Vinjanampathy16,Goold}. However, the laws of thermodynamics have a universal character that applies regardless of whether the system is described by classical or quantum dynamics. For example, entanglement generation cannot help the extraction of work from a quantum system \cite{Hovhannisyan13}, nor in surpassing Carnot efficiency \cite{binder2018}. Nevertheless, thermodynamics does not set bounds on the timescale of such transformations. Indeed, seminal theoretical papers \cite{Binder15,Campaioli17} showed that entangling operations can speed-up the charging process of a quantum battery (QB), a quantum system able to store energy and perform useful work \cite{Alicki13, binder2018}. Inspired by these papers, Ref. \cite{Ferraro17} proposes a quantum Dicke battery, a system where the energy of a photonic cavity mode (acting as a charger) is transferred to a battery consisting of $N$ quantum units described as two-level systems (TLSs). 
Notably, this system displays a collective speed-up of the charging time which decreases as $\sqrt{N}$ \cite{Andolina18c}. 

The Dicke model further exhibits a transitions from quasi-integrability to quantum chaotic dynamics for large light-matter coupling strength \cite{emary2003}, with energy injection into the system enhancing the level of chaos \cite{villasenor2023}. Hence, this chaotic behavior should manifest itself during the charging process. 

The Dicke battery has attracted a great deal of interest given the variety of platforms in which it can be implemented (e.g. superconducting qubits \cite{Wang}, quantum dots \cite{Stockklauser17,Samkharadze18} coupled with a microwave resonator, Rydberg atoms in a cavity \cite{Haroche13}), and numerous variations of this model have been studied
\cite{zhang_arxiv_2018,Zhang18,Crescente20,Crescente20b,Crescente22,Andolina19,Dou22,Dou22b,Zhao}. Recently, a first step towards the realization of a Dicke battery has been experimentally implemented in an excitonic system \cite{Quach}, where a collective boost in the charging process has been reported. 

However, an ideal quantum battery must not only store energy rapidly, but it must be able to provide its stored energy \cite{Andolina19,Monsel19, Maffei21,Tirone}.
In closed quantum systems, the maximum amount of energy that can be extracted from a quantum battery is given by the {\textit{ergotropy}} \cite{Allahvedryan}. When energy is provided to a battery via a quantum charger, correlations between the charger and the battery, and among the units composing the battery, are developed.
Such correlations can greatly limit work extraction \cite{Oppenheim}, and the ergotropy of a single unit of a Dicke battery is very low in standard charging protocols \cite{Andolina19} (later denoted as ``on-off'' protocols).
These detrimental correlations are dramatically larger in chaotic models where entanglement is not limited by a so-called ``are-law'' valid for integrable systems \cite{eisert2010}.

Currently, the development of charging strategies that guarantee a large final ergotropy is still an open problem hindering the usefulness of many-body quantum batteries. Furthermore, while previous literature has often focused on the average energy \cite{Ferraro17,Andolina18c,Andolina19,Crescente20}, in a quantum mechanical setting the energy stored in the battery can fluctuate among each charging instance, leading to a poor {\it charging precision} \cite{Friis18,Rosa20,Delmonte}.

Attempts to maximize the energy stored in a quantum battery have been recently put forward \cite{Mazzoncini,Rodriguez}, where optimal control theory is applied to simple charging scenarios where the charger and the battery are elementary systems, such as a single TLS or a harmonic oscillator. 
However, optimally controlling many-body quantum system, such as the Dicke model, is an extremely challenging task due to the size of the Hilbert space, the chaotic many-body dynamics describing the state evolution, and the difficulty in finding non-analytic control strategies with variational approaches such as Pontryagin's Minimum Principle. For example, in order to optimize the Dicke battery with $N=20$ quantum units, one needs to solve coupled differential equations for more than $4200$ real parameters.

Machine learning techniques, such as Reinforcement Learning (RL) \cite{sutton2018}, have recently proven their strength in tackling complicated optimization problems in a variety of fields, ranging from playing videos games \cite{Christodoulou,Delalleau}, to the board game of GO \cite{Silver}, to controlling plasma \cite{Degrave}. In the field of quantum information and quantum thermodynamics, RL has been used to optimize quantum state preparation \cite{bukov2018,zhang2019,mackeprang2020,brown2021,porotti2022}, error correction \cite{fosel2018,sweke2020}, gate generation \cite{niu2019,an2019,dalgaard2020}, and quantum thermal machines \cite{sgroi2021, ashida2021, Erdman, Erdman1, Erdman2}.

Here we use RL, specifically the soft actor-critic algorithm \cite{haarnoja2018, haarnoja2019}, to discover optimal time-dependent charging protocols for quantum Dicke batteries that overcome the previously described limitation of the standard charging protocol. 
In particular, considering the Dicke quantum battery, including counter-rotating terms, composed of up to $20$ TLSs, we maximize the ergotropy considering two different time-dependent control parameters, i.e. the coupling strength and the frequency detuning between the TLSs and the cavity. Notably, this leads to non-greedy optimal charging strategies that:  (i) provide an ergotropy that almost matches the maximum storable energy, (ii) are fast and can preserve the collective speedup of the charging time, (iii) display a high charging precision, and (iv) do not inject energy through the external controls. 
 This is particularly remarkable given that we modulate a single external control, while dealing with a large Hilbert space whose  dimension scales with the number of units \cite{nielsen1997}.
 
\begin{figure}[!tb]
	\centering
\includegraphics[width=0.9\columnwidth]{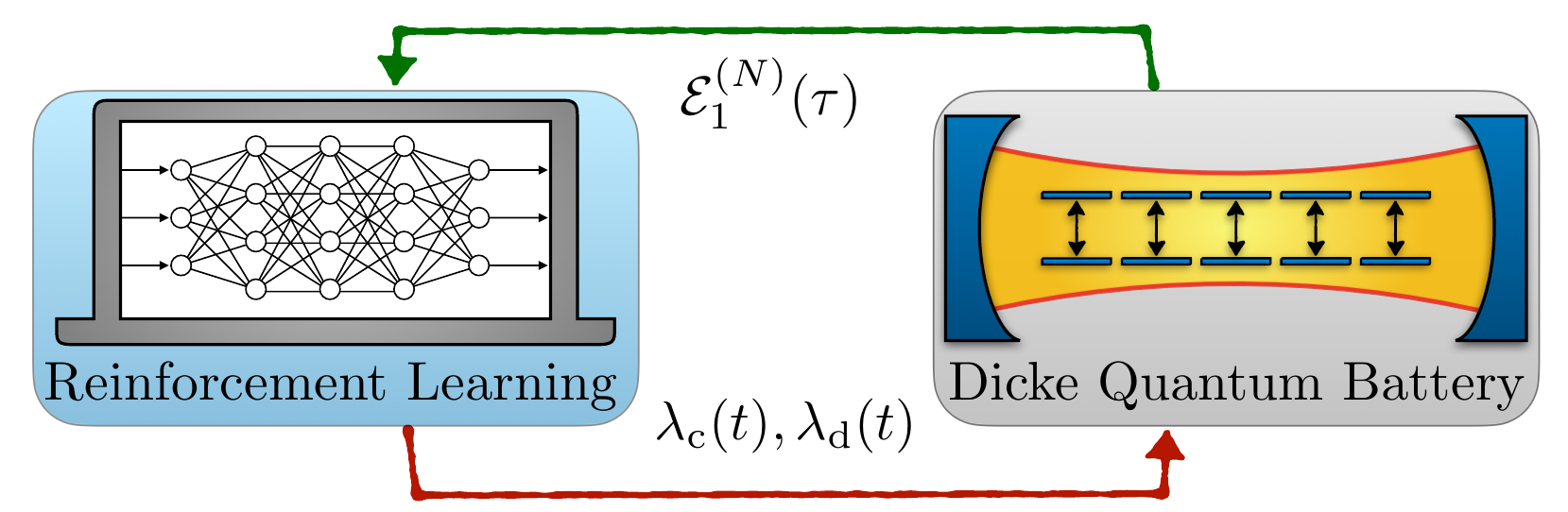}(a)
\includegraphics[width=0.99\columnwidth]{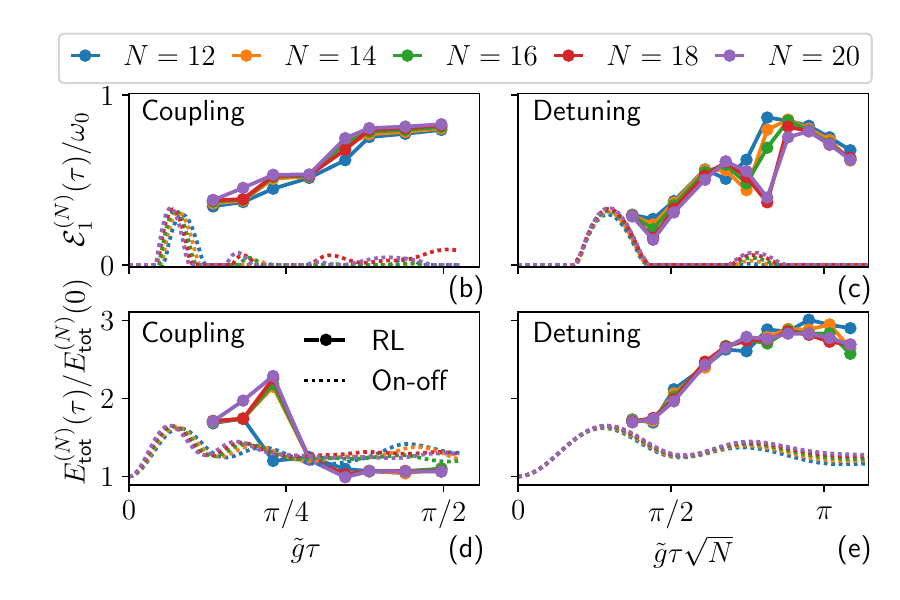}
\caption{(a) A reinforcement learning algorithm maximizes the ergotropy $\mathcal{E}_{1}^{(N)}(\tau)$ of a Dicke quantum battery proposing values of the external control $\lambda_\text{c}(t)$ or $\lambda_\text{d}(t)$ and receiving the variation of ergotropy as reward.
When modulating $\lambda_\text{c}(t)$ (coupling scheme), the
ergotropy $\mathcal{E}^{(N)}_1(\tau)/\omega_0$ and the total energy of the combined charger and battery system
$E^{(N)}_\text{tot}(\tau)/E^{(N)}_\text{tot}(0)$ are plotted respectively in (b) and (d) as a function of the charging time $\tilde{g}\tau$. When modulating $\lambda_\text{d}(t)$ (detuning scheme), $\mathcal{E}^{(N)}_1(\tau)/\omega_0$ and 
$E^{(N)}_\text{tot}(\tau)/E^{(N)}_\text{tot}(0)$ are plotted respectively in (c) and (e) as a function of the rescaled time $\tilde{g}\tau\sqrt{N}$ to show the collective charging speedup.
  Each color in (b-e) corresponds to a different number $N$ of TLSs. A separate RL optimization is performed for each value of $N$ and $\tau$ (large dots), while the dashed lines correspond to the ``on-off'' protocol.
  All optimizations are performed for  $\tilde{g}=0.3\omega_0$ using nearly the same hyperparameters \cite{SM}. In the coupling scheme $\lambda_\text{d}(t)=0$, $\omega_0\lambda_\text{c}(t)\in[-\tilde{g},\tilde{g}]$, and $\Delta t=0.03 \tilde{g}^{-1}$ for $\tau<0.6\tilde{g}^{-1}$, and $\Delta t=0.06\tilde{g}^{-1}$ for $\tau\geq 0.6\tilde{g}^{-1}$. In the detuning scheme $\omega_0\lambda_\text{c}(t)=\tilde{g}$, $\lambda_\text{d}(t)\in[-1,6]$, and $\Delta t=0.11\tilde{g}^{-1}/\sqrt{N}$ to follow the scaling of the charging time.
 }
\label{fig:sketch}
\end{figure}

{\it Protocols and figures of merit.---}
In a Dicke quantum battery, depicted in the gray box of Fig.~\ref{fig:sketch}(a), energy is stored in $N$ TLSs each corresponding to a single unit of the battery. When the battery is isolated, the TLSs are governed by the following free and local battery Hamiltonian ($\hbar = 1$),
$ \mathcal{\hat H}_{\rm B} = \sum_{j=1}^N \hat{h}_j^{\rm B}~$,
where $\hat{h}_j^{\rm B}=\omega_0/2\big(\hat{\sigma}^{(z)}_j+1\big)$, $\omega_0$ is the energy splitting between the excited state $\ket{1}_j$ and the ground state $\ket{0}_j$ of the TLS, and $\hat{\sigma}^{(\alpha)}_j$ are the $\alpha=x,y,z$ Pauli matrices acting on the $j-$th TLS.
Energy is provided by a charger, a single mode cavity resonant with the TLSs at frequency $\omega_0$, described by $\mathcal{\hat H}_{\rm C} =\omega_0\hat{a}^\dagger\hat{a}~$, where $\hat{a}^\dagger,\hat{a}$ are the bosonic ladder operators. 
At time $t=0$, the battery starts interacting with the charger. The initial state is assumed to be the tensor product of the TLSs' ground states, $\ket{\rm G}\equiv\otimes_{j=1}^N\ket{0}_j$, physically representing the discharged battery, while the cavity is assumed to be in an $N$ photon Fock state $\ket{N}$, hence  $\ket{\psi(0)}=\ket{{\rm G}}\otimes\ket{N}$, $\ket{\psi(t)}$ being the total wave-function. Given the resonant condition, the energy in the charger is exactly enough to potentially fully charge the battery.
The system then evolves according to the time-dependent Schr{\"o}dinger equation $i\partial_t\ket{\psi(t)}=\mathcal{\hat H}(t)\ket{\psi(t)}$ where \cite{Andolina18}
\begin{equation}\label{eq:H_tot}
\mathcal{\hat H}(t) = \mathcal{\hat H}_{\rm C}+(1+\lambda_\text{d}(t))\,\mathcal{\hat H}_{\rm B}+\lambda_\text{c}(t)\,\mathcal{\hat H}_{\rm int}~,
\end{equation}
$\mathcal{\hat H}_{\rm int}=\omega_0\sum_{j=1}^N\hat{\sigma}^{(x)}_j (\hat{a}+\hat{a}^\dagger)$ is the charger-battery interaction Hamiltonian, and $\lambda_\text{c}(t)$, $\lambda_\text{d}(t)$ are classical external control parameter determining, respectively, the coupling strength, and the detuning of the TLSs. After time $\tau$, dubbed the {\it charging time}, the external parameters are switched off, i.e. $\lambda_\text{c}(t)=\lambda_\text{d}(t)=0$, which corresponds to decoupling the battery from the charger and to removing the detuning. 
Notice that the interaction term in Eq.~(\ref{eq:H_tot}) differs from some literature by a factor $\sqrt{N}$, such that the model becomes chaotic for $\lambda_\text{c}(t)\sqrt N > 1/4$ \cite{emary2003}. We study the system in the chaotic regime, where the counter-rotating terms in $\mathcal{\hat H}_{\rm int}$ cannot be neglected \cite{SM}. 

We consider two charging schemes: in the \textit{coupling scheme}, we modulate the coupling strength $\lambda_\text{c}(t)$ without any detuning ($\lambda_\text{d}(t)=0$). In the \textit{detuning scheme} we fix $\lambda_\text{c}(t)$ to a constant, and we only modulate the detuning $\lambda_\text{d}(t)$. We then compare these to the commonly employed ``on-off'' charging protocol \cite{Ferraro17, Le17, Andolina19}, which corresponds to setting $\lambda_\text{c}(t)=\tilde{g}/\omega_0$ and $\lambda_\text{d}(t)=0$ for $t\in [0,\tau]$, where $\tilde{g}$ represents the largest effective coupling strength.

The mean energy stored in the battery at the end of the protocol is given by $E^{(N)}(\tau) =
\braket{\psi(\tau) |\mathcal{\hat H}_{\rm B}| \psi(\tau)}$. However, not all of the energy $E^{(N)}(\tau)$ can be extracted; indeed, interactions with the cavity can create correlations between the cavity and the battery, and  between the units of the battery, thus deteriorating the extractable work \cite{Andolina19}. The energy that can be extracted from a single battery unit is given by the ergotropy of the TLS \cite{Farina18}
\begin{equation}\label{eq:Ergo}
\mathcal{E}^{(N)}_1(\tau)=\frac{E^{(N)}(\tau)}{N}-r_{1}(\tau)\omega_0~,
\end{equation}
 where $r_{1}(\tau)$ is the minimum eigenvalue of the single TLS reduced density matrix ${\rho}_{{\rm B},1}(\tau)$. Details on the calculation of the ergotropy are given in the SM \cite{SM}.
 
Undesired energy can be injected into the system by the modulation of the external controls. We quantify this analyzing the variation of the total energy of the combined charger-battery system $E^{(N)}_\text{tot}(\tau) = \braket{\psi(\tau) |\mathcal{\hat H}(\tau)| \psi(\tau)}$. 
We further quantify the charging precision at the end of the charging protocol computing the variance of the energy stored in a single battery unit: \begin{equation}
    \sigma^2_{E_1^{(N)}}(\tau) = 
    \braket{\psi(\tau) |(\hat h^\text{B}_{1})^2| \psi(\tau)} - \braket{\psi(\tau) |\hat h^\text{B}_{1}| \psi(\tau)}^2.
\end{equation}

{\it Results and discussion.---}
We use RL to maximize the ergotropy $\mathcal{E}^{(N)}_1(\tau)$ for various charging times $\tau$.
Discretizing time in steps of duration $\Delta t$ during which the controls are constant, the RL method determines the values of $\lambda_\text{c}(t)$ or $\lambda_\text{d}(t)$ that maximizes the final ergotropy $\mathcal{E}^{(N)}_1(\tau)$ (see SM \cite{SM} for details on the RL method).

Figure~\ref{fig:sketch}(b,c) reports the optimized single battery ergotropy $\mathcal{E}^{(N)}_1(\tau)$ using the coupling and detuning schemes respectively. Each dot along the full lines represents a separate optimization using RL for different values of the battery size $N$ (each one corresponding to a different color), and for different charging times $\tau$ reported on the x-axis, whereas the dotted lines correspond to the ``on-off'' strategy. 
The RL optimization is not reported for small $\tau$, as it coincides with the ``on-off'' strategy until the peak of the ergotropy is reached.

First, we notice that the charging protocols discovered with RL substantially outperform the ``on-off'' protocol. Indeed, while ``on-off'' protocols initially reach an ergotropy of $\sim 30 \%$ of $\omega_0$ and then essentially decay to zero - the ergotropy delivered by the RL protocols reaches roughly $87\%$ of $\omega_0$, corresponding to almost full energy extraction from the nearly fully charged battery. However, this comes at the expense of an increased charging time $\tau$.

\begin{figure}[!tb]
	\centering
	\includegraphics[width=0.99\columnwidth]{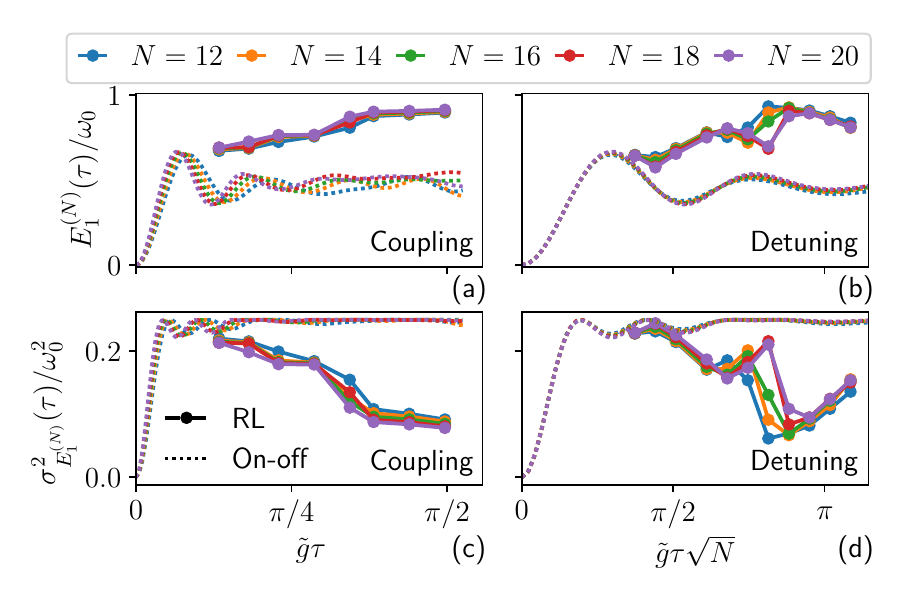}
	\caption{The energy $E_1^{(N)}(\tau)/\omega_0$ stored in a single battery unit is plotted as a function of $\tilde{g}\tau$ in the coupling scheme (a), and as a function of $\tilde{g}\tau\sqrt{N}$ in the detuning scheme (b). The corresponding energy variance $\sigma^2_{E_1^{(N)}}(\tau)/\omega_0^2$ is displayed in (c,d). These plots correspond to the results presented in Fig.~\ref{fig:sketch}(b-e) using the same color-code and line style. }
	\label{fig:energy}
\end{figure}
\begin{figure*}[!t]
    \centering
    \includegraphics[width=\textwidth]{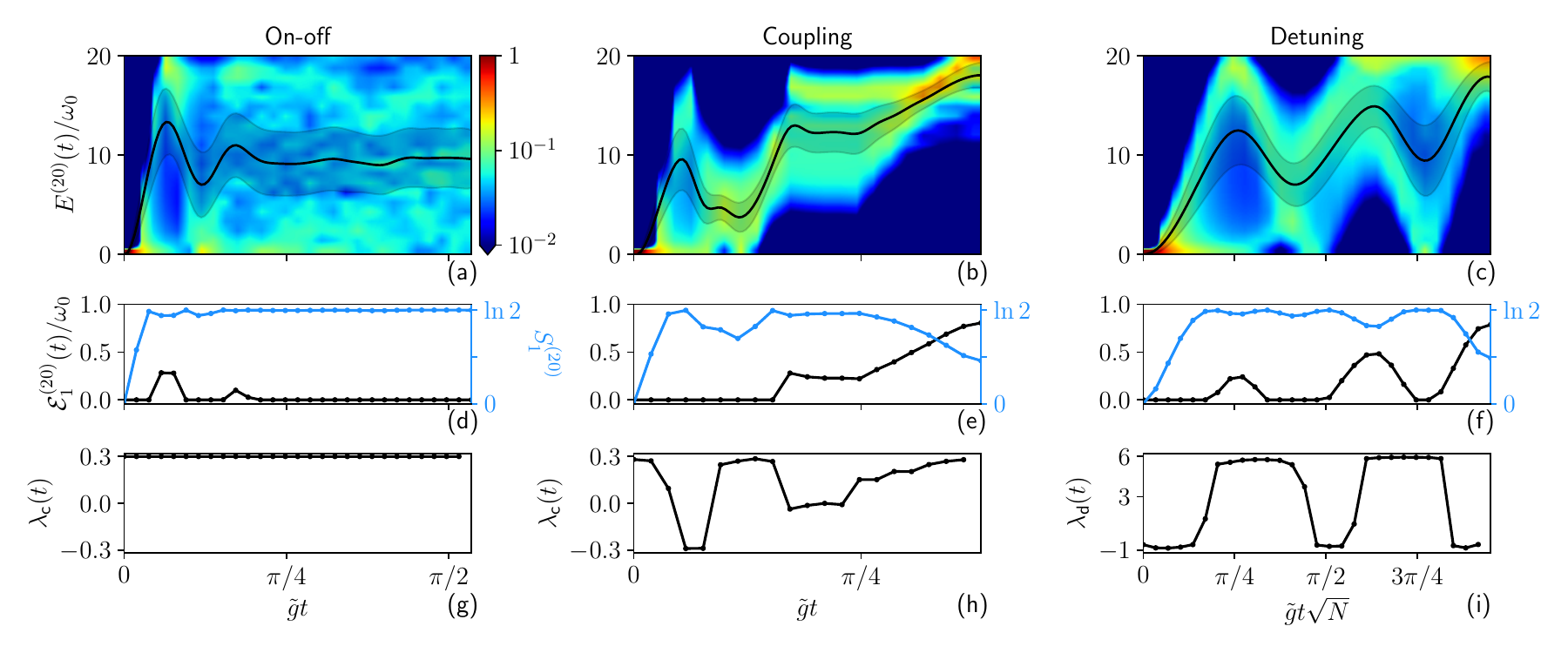}
    \caption{Performance of the on-off (left), coupling (middle) and detuning (right) cases as a function of time. (a-c): Density plot of the squared projection of $\ket{\psi(t)}$ onto the spectrum of $\hat{\mathcal{H}}_\text{B}$ (the energy of the spectrum is on the y-axis). The average $E^{(20)}(t)$ and standard deviation of the TLSs' energy is shown as a black line and corresponding shadowed area. (d-f): Single TLS ergotropy $\mathcal{E}_1^{(20)}(t)$ (black curve) and entropy $S_1^{(20)}$ (blue curve). (g-i): Corresponding values of the control. Each small dot in (d-e) corresponds to a time-step $\Delta t$ in the RL approach. The data corresponds to the optimization carried out in Fig.~\ref{fig:sketch} choosing $N=20$, $\tilde{g}\tau=1.68$ in the on-off case, $\tilde{g}\tau=1.2$ in the coupling case, and $\tilde{g}\tau\sqrt{N}=2.99$ in the detuning case.
}
    \label{fig:energy_density}
\end{figure*}
Most notably, we find evidence that the RL charging protocols in the detuning scheme preserve the collective speed-up of the charging power. Indeed, the charging curves in Figs.~\ref{fig:sketch}(b,c) overlap for different values of $N$ only when plotted as a function of $\tilde{g}\tau$ in the coupling case, and as a function of $\tilde{g}\tau\sqrt{N}$ in the detuning case. This suggests that the charging time $\tau$ decreases as $1/\sqrt{N}$ in the detuning case, thus extending the collective speed-up, originally found for ``on-off'' protocols in \cite{Ferraro17}, to values of the ergotropy close to its theoretical maximum. Notice that, to highlight this effect, we kept $\Delta t$ constant in the coupling case, while we scale it as $\sim 1/\sqrt{N}$ in the detuning case
(see SM for additional details, and Fig. S1 therein for the equivalent of Fig.~\ref{fig:sketch} with inverted scaling of the charging time).

In Fig.~\ref{fig:sketch}(d,e) we report the corresponding energy injected into the system by the modulation of the controls. Interestingly, in the coupling scheme $E^{(N)}_\text{tot}(\tau)/E^{(N)}_\text{tot}(0)$ reaches $1$ for large $\tau$, which corresponds to no external energy injection, performing better than the ``on-off'' strategy that injects energy into the system even at null ergotropy. In the detuning case, however, the energy injection is high and reaches up to $E^{(N)}_\text{tot}(\tau)/E^{(N)}_\text{tot}(0)\approx 3$. This seems to highlight the existence of a trade-off between injected energy, and collective charging speed-up, which could be interpreted as a manifestation
of the Margolus-Levitin quantum speed limit \cite{margolus1998}.

In Fig.~\ref{fig:energy}(a,b) we plot the energy stored in a single battery unit $E_1^{(N)}(\tau)$ respectively in the coupling and detuning cases, confirming the scaling of the charging time of the two schemes. As expected from the high values of the ergotropy, we see that the RL-discovered protocols nearly reach full charge, while the ``on-off'' protocol only reaches $\sim 50\%$ of $\omega_0$. In Fig.~\ref{fig:energy}(c,d) we see that the RL protocols simultaneously enhance also the charging precision. Indeed, the variance $\sigma^2_{E_1^{(N)}}(\tau)$ roughly decreases with increasing $\tau$, whereas it remains constant at a maximum value in the on-off case.
 
We now investigate the origin of the performance boost found with RL from the point of view of quantum chaos. 
The performance of the on-off protocol is limited because we are in the chaotic regime. Indeed, local sub-systems of a quantum chaotic model are highly entangled to the rest of the system, therefore they are in a nearly thermal state. Since a thermal state is passive \cite{Allahvedryan}, hardly any energy can be extracted from a battery unit \cite{Rossini19}. 
We show that RL learns to counter the detrimental effect of quantum chaos by (i) focusing the state onto high energy eigenstates of the battery Hamiltonian, and (ii) leading to an inversion of the natural increase of the local entropy of the individual TLSs, which measures the correlations to the rest of the system.

In Figs.~\ref{fig:energy_density}(a-c) we display a density plot of the squared projection of $\ket{\psi(t)}$ onto the spectrum of $\hat{\mathcal{H}}_\text{B}$ for $N=20$ (other values of $N$ are qualitatively similar), for a value of $\tau$ that leads to large final ergotropy $\mathcal{E}_1^{(20)}(\tau)$. The black curve and region represent respectively the average and standard deviation of the energy of the TLSs. In the on-off case, Fig.~\ref{fig:energy_density}(a), after the first oscillation, the state quickly spreads onto a roughly uniform distribution of all eigenstates. The manifestation of chaos is even more clear in the single battery entropy $S^{(20)}_1=-\text{Tr}[\rho_{\text{B},1}(t)\ln\rho_{\text{B},1}(t)]$, shown as a blue curve in Fig.~\ref{fig:energy_density}(d), which quickly reaches and plateaus to the maximum value $\ln 2$, corresponding to a high temperature thermal state. Therefore the ergotropy [black curve in Fig.~\ref{fig:energy_density}(d)] drops to zero.

A stark difference is visible when comparing to Figs.~\ref{fig:energy_density}(b,c), where RL is able to counter the onset of chaos by (i) squeezing the energy distribution around the highest energy eigenstates at the final time $\tau$ (red region in the upper right), and  (ii) reducing the entropy $S^{(20)}_1$ [blue curve in Fig.~\ref{fig:energy_density}(e,f)], which in turn leads to a rapid increase of the ergotropy [black curve in Fig.~\ref{fig:energy_density}(e,f)]. This is achieved thanks to the oscillatory charging protocol reported in Fig.~\ref{fig:energy_density}(h,i) \cite{SM}.

This can be intuitively understood in the coupling case. In the interaction picture, the dynamics is governed by the interaction picture Hamiltonian $\lambda_\text{c}(t)\tilde{\mathcal{H}}_\text{int}(t)$ \cite{SM}.  When we switch the sign of the control $\lambda_\text{c}(t)$ [see Fig.~\ref{fig:energy_density}(h)], we are changing the sign of the interaction which, for short times, approximately inverts the arrow of time, thus the entropy \cite{huang1963}. However, the exact optimal modulation of the control is far from trivial. A similar effect is observed in spin echo, where a laser pulse is used to invert the dynamics of $N$ spins, hence countering the detrimental effect of dephasing \cite{hahn1950}.

We finally notice that the non-monotonic behavior of the energy and ergotropy in Fig.~\ref{fig:energy_density}(e,f) denotes that the RL charging protocol is non-greedy, i.e. it learns to sacrifice the ergotropy for short times to reach a higher final ergotropy at time $\tau$ (see SM \cite{SM} for details).

{\it Conclusions}.— We employed reinforcement learning to discover optimal charging protocols for a Dicke many-body battery, composed of up to $N=20$ units, either modulating the coupling strength or the detuning of the TLSs.
Using the standard ``on-off'' charging strategy, the ergotropy of a single battery unit does not exceed $\sim30\%$ of the total energy, exhibits a low charging precision, and energy is externally injected into the system through the coupling modulation.
Using RL, we can simultaneously boost the ergotropy up to $87\%$ of the maximum storable energy, and enhance the charging precision reducing quantum fluctuations by more than $50\%$, and we interpret these results from the point of view of quantum chaos. Notably, in the detuning scheme, we find evidence that the collective speedup of the charging time with increasing $N$ can be preserved even when nearly fully charging the battery. Conversely, in the coupling scheme, we can nearly fully charge the battery without injecting any external energy. This points to the existence of a trade-off between collective speedups in the charging speed, and reducing external energy injection.
Interestingly, we find that optimal charging strategies are non-greedy and $\tau$-dependent, i.e., when $\tau$ is large, the RL method learns to sacrifice ergotropy at short times, to reach a higher ergotropy at the final time $\tau$.
These results could be tested directly in experimental platforms such as the $10$ superconducting qubit device of Ref. \cite{Wang}. 

In the future, the present RL method could be fruitfully used to optimize the charging process of numerous other many-body batteries, such as spin-chains batteries \cite{Le17,Rossini19,Ghosh19,Farre18} and Sachdev–Ye–Kitaev batteries \cite{Rossini20}, the latter both saturating the power bound \cite{Campaioli17,Gyhm22} and displaying strongly chaotic dynamics. The effect of dissipation during the charging could also be considered \cite{Farina18,Barra19,Pirmoradian19,Quach20,Gherardini19,Seah,Salvia,Shaghaghi,Liu19} and feedback control strategies can be investigated \cite{Gherardini19,Mitchison21}. 

The RL code is publicly available \cite{github}.
Numerical work has been performed by using PyTorch \cite{paszke2019} and QuTiP2 toolbox \cite{QuTip}.

{\it Acknowledgments}.— We
wish to thank M. Polini, D. Ferraro, V. Cavina, F. Campaioli and G. Calaj{\`o} for useful discussions. F.N gratefully acknowledges funding by the BMBF (Berlin Institute for the Foundations of Learning and Data -- BIFOLD), the European Research Commission (ERC CoG 772230) and the Berlin Mathematics Center MATH+ (AA1-6, AA2-8). P.A.E gratefully acknowledges funding by the Berlin Mathematics Center MATH+ (AA1-6, AA2-18). V.G. acknowledge financial support by MIUR (Ministero dell’ Istruzione, dell’ Universit{\`a} e della Ricerca) by PRIN 2017 Taming complexity via Quantum Strategies: a Hybrid Integrated Photonic approach (QUSHIP) Id. 2017SRN-BRK, and via project PRO3 Quantum Pathfinder. 
G.M.A. acknowledges funding from the European Research Council (ERC) under the European Union’s Horizon 2020 research and innovation programme (Grant agreement No. 101002955 — CONQUER).

\clearpage 
\setcounter{section}{0}
\setcounter{equation}{0}%
\setcounter{figure}{0}%
\setcounter{table}{0}%

\setcounter{page}{1}

\renewcommand{\thetable}{S\arabic{table}}
\renewcommand{\theequation}{S\arabic{equation}}
\renewcommand{\thefigure}{S\arabic{figure}}

\onecolumngrid

\begin{center}
\textbf{\large Supplemental Material for:\\ ``Reinforcement learning optimization of the charging of a Dicke quantum battery''}
\bigskip

Paolo Andrea Erdman, $^{1,\,*}$
Gian Marcello Andolina,$^{2,\,3,\,*}$
Vittorio Giovannetti,$^{4}$ and
Frank No{\'e}$^{5,\,1,\,6,\,7}$

\bigskip
$^1$\!{\it Freie Universit{\" a}t Berlin, Department of Mathematics and Computer Science, Arnimallee 6, 14195 Berlin, Germany}

$^2$\!{\it ICFO-Institut de Ci\`{e}ncies Fot\`{o}niques, The Barcelona Institute of Science and Technology, Av. Carl Friedrich Gauss 3, 08860 Castelldefels (Barcelona),~Spain}

$^3$\!{\it JEIP, UAR 3573 CNRS, Coll\`ege de France, PSL Research University, F-75321 Paris, France}

$^4$\!{\it NEST, Scuola Normale Superiore and Istituto Nanoscienze-CNR, I-56126 Pisa,~Italy}

$^5$\!{\it Microsoft Research AI4Science, Karl-Liebknecht Str. 32, 10178 Berlin, Germany}

$^6$\!{\it Freie Universit{\" a}t Berlin, Department of Physics, Arnimallee 6, 14195 Berlin, Germany}

$^7$\!{\it Rice University, Department of Chemistry, Houston, TX 77005, USA}

$^*$\!{ These two authors contributed equally.}

\end{center}

\bigskip

In this Supplemental Material we provide additional information on the importance of the counter-rotating terms, on the calculation of the ergotropy, on the scaling of the charging time, 
on the non-greedy nature of RL charging protocols, we explicitly show additional charging protocols and we provide details on the Reinforcement Learning (RL) method.

\appendix

\section{Importance of the counter-rotating terms}
\label{Appendix:Details1}
Here we discuss the importance of the counter-rotating terms, showing that it is not possible to neglect them, even in the weak coupling regime.
The Dicke battery Hamiltonian in terms of collective operators, $\hat{J}_z=\sum_{j=1}^N\hat{\sigma}_j^{(z)}/2$ and $\hat{J}_\pm=\sum_{j=1}^N\hat{\sigma}_j^{\pm} $, is given by the following Hamiltonians,
\begin{eqnarray}
\hat{\mathcal{H}}_{\rm C}&=&\omega_0\hat{a}^\dagger \hat{a}~,\\
\hat{\mathcal{H}}_{\rm B}&=&\omega_0(\hat{J}_z+\frac{N}{2})~,\\
{\mathcal{\hat H}}_{\rm int}&=&2\omega_0(\hat{J}_+ +\hat{J}_-) (\hat{a}+\hat{a}^\dagger).
\end{eqnarray}

It is useful to rewrite the interaction Hamiltonian in the interaction picture, $\tilde{\mathcal{H}}_{\rm int}(t)\equiv e^{i\hat{\mathcal{H}}_0t} \hat{\mathcal{H}}_{
\rm int} e^{-i\hat{\mathcal{H}}_0t}$, where $\hat{\mathcal{H}}_0=\hat{\mathcal{H}}_{\rm B}+\hat{\mathcal{H}}_{\rm C}$. We have
\begin{eqnarray}
\label{eq:H_intpic}
\tilde{\mathcal{H}}_{\rm int}(t)=2\omega_0(\hat{J}_+e^{i\omega_0t} +\hat{J}_- e^{-i\omega_0t})(\hat{a}e^{-i\omega_0t}+\hat{a}^\dagger e^{i\omega_0t}).
\end{eqnarray}
 We remind that in this reference frame the dynamics is dictated by the interaction Hamiltonian only and thus the wave-function at time $\tau$ is given by $\ket{\psi(\tau)}=\hat{\mathcal{T}} \exp[-i\int_0^\tau  \lambda(t)\tilde{\mathcal{H}}_{\rm int}(t) dt]\ket{\psi_0}$, where $\hat{\mathcal{T}}$ is the time-ordering operator. In the weak coupling regime, i.e. when $\lambda_\text{c}(t)$ is constant and $\lambda_\text{c}(t) \ll 1$, it is customary to neglect fast-oscillating counter-rotating terms, $e^{i2\omega_0t}\hat{J}_+\hat{a}^\dagger,e^{-i2\omega_0t}\hat{J}_-\hat{a}$ in Eq.~\eqref{eq:H_intpic}. Nevertheless, in the case under study it is not possible to perform this approximation, since $\lambda_\text{c}(t)$ can oscillate at frequency $\pm2\omega_0$ and compensate for the fast oscillations.

\section{Details on ergotropy calculation}
\setcounter{equation}{4}%

\label{app:ergotropy}

In this section we detail the calculation of the ergotropy of a single TLS, deriving Eq. (2) of the main text. 
The maximum amount of energy, measured with respect to a local Hamiltonian $\mathcal H$, that can be extracted from a quantum state $\rho$ by using arbitrary unitary transformations is given by the ergotropy $\mathcal{E}(\rho,{\hat{\mathcal H}})$. A closed expression for this quantity is given by the difference 

\begin{equation}\label{eq:ergotropy2}
\mathcal{E}(\rho,\hat{\cal H})=E(\rho)-E(\tilde{\rho})\;
\end{equation}
between the mean energy  $E(\rho) = {\rm tr}[\hat{\mathcal{H}} \rho]$  of the state $\rho$ and of the mean energy  $E(\tilde{\rho}) = {\rm tr}[\hat{\mathcal{H}}  \tilde{\rho}]$ of the passive state $\tilde{\rho}$ associated with $\rho$. The latter is defined as the density matrix which is diagonal on the eigenbasis of $\hat{\mathcal{H}} $ and whose eigenvalues correspond to a proper reordering of those of $\rho$, i.e.   $\tilde{\rho}=\sum_n r_n \ket{\epsilon_n}\bra{\epsilon_n}$ with 
$\rho=\sum_n r_n\ket{r_n}\bra{r_n} $, $\hat{\mathcal{H}} =\sum_n \epsilon_n\ket{\epsilon_n}\bra{\epsilon_n}$, with  $r_0\geq r_1 \geq \cdots$ and $\epsilon_0\leq \epsilon_1 \leq \cdots$, yielding
\begin{equation}
\label{passive}
E(\tilde{\rho})=\sum_{n} r_n \epsilon_n~.
\end{equation}
In the problem at hand, we focus on the ergotropy of a single battery unit, consisting of a TLS.
In this case, the density matrix at time $\tau$ is given by the $2\times 2$ matrix $\rho_{{\rm B},1}(\tau)$, while the energy is measured with respect to the Hamiltonian $\hat{h}_1^{\rm B}=\omega_0\big(\hat{\sigma}^{(z)}_1+1/2\big)$. Here, we can chose the first TLS, $j=1$, without any loss of generality, due to the invariance under TLS permutations of the Dicke Hamiltonian. Thus, the energy that can be extracted from a single battery unit reads
\begin{eqnarray}
\mathcal{E}_{1}^{(N)}(\tau) \equiv \mathcal{E}(\rho_{{\rm B},1}(\tau),\hat{h}_1^{\rm B})~. \label{ERGO}
\end{eqnarray}
 This expression can be further simplified by expressing the $\rho_{{\rm B},1}(\tau)$ in a diagonal basis,
\begin{eqnarray}
\rho_{{\rm B},1}(\tau)= r_{0}(\tau)\ket{r_0(\tau)}\bra{r_0(\tau)}+r_{1}(\tau)\ket{r_1(\tau)}\bra{r_1(\tau)}~, \label{rho_diag}
\end{eqnarray}
where the eigenvalue are ordered such that $r_0 (\tau)\geq r_1(\tau)$. In this case, the ergotropy $\mathcal{E}_{1}^{(N)}(\tau)$ simplifies to
\begin{eqnarray}
\mathcal{E}_{1}^{(N)}(\tau) =\frac{E^{(N)}(\tau)}{N}-r_{1}(\tau)\omega_0~, \label{ERGOS}
\end{eqnarray}
where we used that ${\rm tr}[\rho_{{\rm B},1}(\tau)\hat{h}_1^{\rm B}]=({E^{(N)}(\tau)}/{N})$ due to permutation symmetry.
The Dicke Hamiltonian (cf. Eq.~(1) in the main text) can be rewritten in terms of collective operators $\hat{J}_\alpha=\sum_{j=1}^N\hat{\sigma}_j^{(\alpha)}/2$ with $\alpha=x,y,z$. The numerical calculations have been performed in the so-called Dicke basis, where states are described by the total and the z- angular momentum, $\hat{J}^2=\sum_{\alpha}\hat{J}^2_\alpha$ and $\hat{J}_z$. In this basis, the battery density matrix ${\rho}_{\rm B}$ can be written in the Dicke basis as follows,
\begin{eqnarray}\label{rho}
\rho_{\rm B}=\sum_{J,M,J^\prime,M^\prime}\rho_{J,M,J^\prime,M^\prime} \ket{J,M}\bra{J^\prime,M^\prime}~,
\end{eqnarray}
 $J,M$ being the eigenvalues associated with the total and the $z$- angular momentum (Notice that we dropped the dependence upon the charging time $\tau$ for the sake of conciseness). Since the eigenvalue $J$ associated with the total angular momentum $\hat{J}^2=J(J+1)$ is a well defined quantum number and the initial state of the system is given by the ground state $\ket{J=N/2,M=-N/2}=\ket{\rm G}$, the dynamics is restricted to the manifold $J=N/2$,
\begin{eqnarray}\label{rhob}
\rho_{\rm B}=\sum_{M,M^\prime}\rho_{N/2,M,N/2,M^\prime} \ket{N/2,M}\bra{N/2,M^\prime}~.
\end{eqnarray}
 We now express the density matrix in the uncoupled basis $\ket{s_1,\dots,s_N}=\otimes_{i=1}^N\ket{s_i}_i$, where $s_i=0$ ($s_i=1$) denotes the $i$-th atoms being in the ground (excited) state,
\begin{eqnarray}\label{rho1}
 {\rho}_{\rm B}=\sum_{M,M^\prime} \sum_{s_1,\dots, s_N} \sum_{s_1^\prime,\dots, s^\prime_N}\ket{s_1,\dots, s_N} \braket{s_1,\dots, s_N|N/2,M}\rho_{N/2,M,N/2,M^\prime} \braket{N/2,M^\prime|s^\prime_1,\dots, s^\prime_N} \bra{s^\prime_1,\dots, s^\prime_N}~.\nonumber \\
\end{eqnarray}
The scalar product $\braket{N/2,M^\prime|s^\prime_1,\dots, s^\prime_N}$ can be calculated recalling that $\ket{N/2,M^\prime}$ expressed in terms of $s^\prime_1,\dots, s^\prime_N$ is given by the completely symmetric combination
\begin{eqnarray}\label{scalar}
\ket{N/2,M}=\sum_{s^\prime_1,\dots, s^\prime_N}\binom{N}{\frac{N}{2}+M}^{-\frac{1}{2}} \delta_{M,M(\{s_i\})}\ket{s^\prime_1,\dots, s^\prime_N}~,
\end{eqnarray}
where $M(\{s_i\})=N/2-\sum_{i=1}^N s_i$ and the exact pre-factor has been obtained by imposing the normalization of the wave-function.
Hence the overlap $\braket{N/2,M^\prime|s^\prime_1,\dots, s^\prime_N}$ reads
\begin{eqnarray}\label{scalar1}
\braket{N/2,M|s_1,\dots, s_N}=\binom{N}{\frac{N}{2}+M}^{-\frac{1}{2}} \delta_{M,M(\{s_i\})}~.
\end{eqnarray}
 The previous expression shows that the z- angular momentum is fully determined by the number of excitations in the systems. Hence we have
\begin{eqnarray}\label{rho2}
 {\rho}_{\rm B}=  \sum_{s_1,\dots, s_N} \sum_{s_1^\prime,\dots, s^\prime_N}\rho_{N/2,M(\{s_i\}),N/2,M(\{s^\prime_i\})} \binom{N}{\frac{N}{2}+M(\{s_i\})}^{-\frac{1}{2}}\binom{N}{\frac{N}{2}+M(\{s^\prime_i\})}^{-\frac{1}{2}} \ket{s_1,\dots, s_N}  \bra{s^\prime_1,\dots, s^\prime_N}~.\nonumber \\
\end{eqnarray}
 We are interested in the density matrix of the first TLS, obtained tracing out all other TLSs, $\rho_{{\rm B},1}={\rm tr}_{s_2,\dots,s_N} [{\rho}_{{\rm B}}]$, which reads
 
 \begin{eqnarray}\label{rho3}
\rho_{{\rm B},1}= \sum_{s_2,\dots, s_N}\rho_{N/2,M(\{s_i\}),N/2,M(\{s^\prime_i\})} \binom{N}{\frac{N}{2}+M(\{s_i\})}^{-\frac{1}{2}}\binom{N}{\frac{N}{2}+M(\{s^\prime_i\})}^{-\frac{1}{2}} \ket{s_1}  \bra{s^\prime_1}~.\nonumber \\
\end{eqnarray}
It is useful to define the number of excitations in the other TLSs, $e=\sum_{i=2}^N s_i$. We note that the expression of the density matrix in Eq. \eqref{rho3} does not depends on the specific values $s_2,\dots, s_N$ but only over the sum of all values, which is given by $e$. Hence we can sum only over the variables $e$, as follows

 \begin{eqnarray}\label{rho4}
\rho_{{\rm B},1}= \sum_{e=0}^{N-1}\rho_{N/2,e+s_1-N/2,N/2,e+s^\prime_1-N/2} \binom{N-1}{e} \binom{N}{e+s_1}^{-\frac{1}{2}}\binom{N}{e+s^\prime_1}^{-\frac{1}{2}} \ket{s_1}  \bra{s^\prime_1}~,\nonumber \\
\end{eqnarray}
where the factor $\binom{N-1}{e}$ takes into account the degeneracy of states with different $s_2,\dots, s_N$ but same number of excitations $e$. This expression can be further simplified as,

 \begin{eqnarray}\label{rho5}
\rho_{{\rm B},1}= \sum_{e=0}^{N-1}\rho_{N/2,e+s_1-N/2,N/2,e+s^\prime_1-N/2} \frac{1}{N} \frac{\sqrt{(e+s_1)!(e+s^\prime_1)!}}{e!} \frac{\sqrt{(N-s_1-e)!(N-s^\prime_1-e)!}}{(N-1-e)!} \ket{s_1}  \bra{s^\prime_1}~.\nonumber \\
\end{eqnarray}
The previous equation gives the density matrix $\rho_{{\rm B},1}$. Thus it is sufficient to diagonalize it and use Eq.~\eqref{ERGOS} to obtain the ergotropy of a TLS.

\section{Scaling of the charging time}
\begin{figure}[htbp]
    \centering
    \begin{minipage}{0.47\textwidth}
        \centering
        \includegraphics[width=\linewidth]{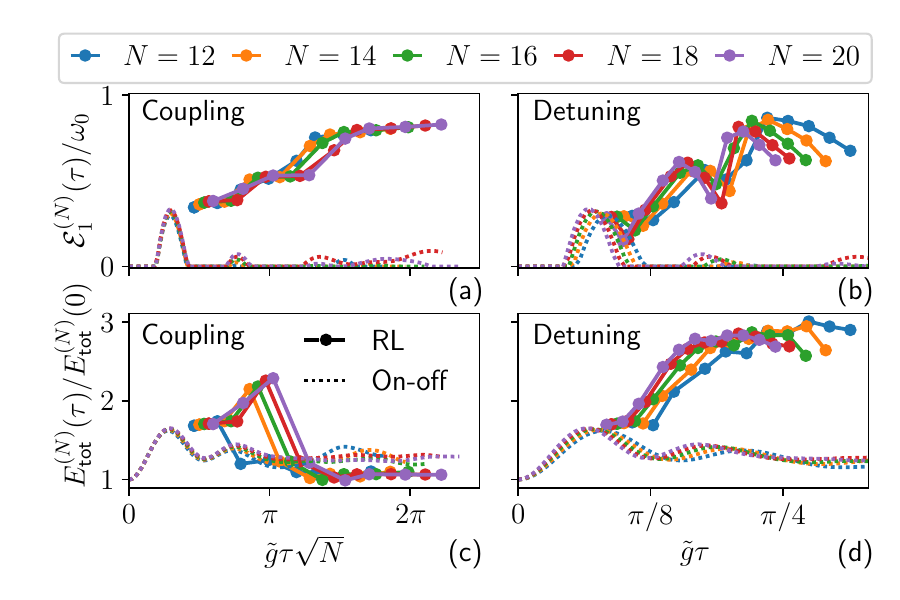}
        \caption{Same plot as Fig.~1(b-e) of the main text, but with inverted scaling of charging time $\tau$ on the x-axis. More specifically, in the coupling scheme the ergotropy and total energy of the charger and battery system are plotted as a function of $\tilde{g}\tau\sqrt{N}$, while they are plotted as a function of $\tilde{g}\tau$ in the detuning case.}
        \label{fig:ergo_inv_time}
    \end{minipage}
    \hfill
    \begin{minipage}{0.47\textwidth}
        \centering
        \includegraphics[width=\linewidth]{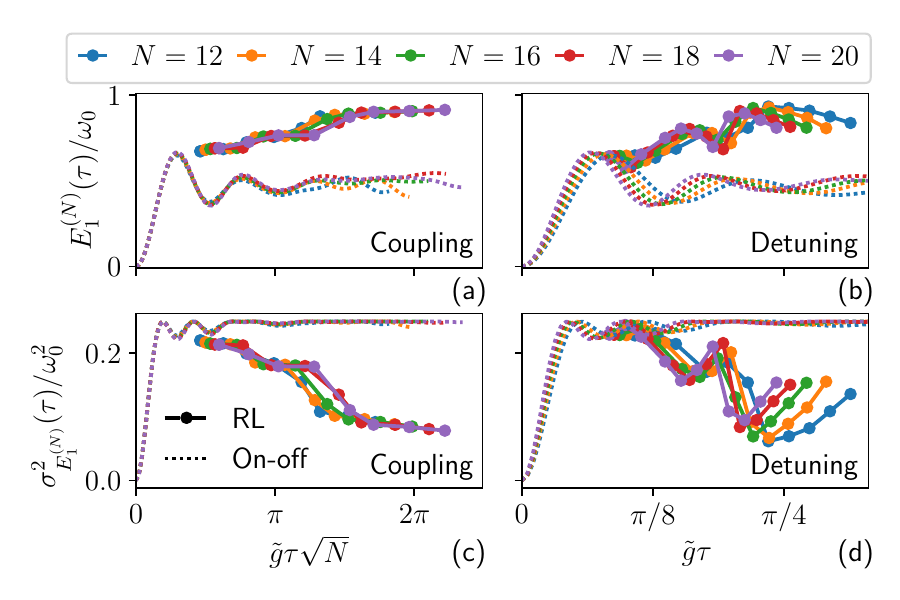}
        \caption{Same plot as Fig.~2 of the main text, but with inverted scaling of charging time $\tau$ on the x-axis. More specifically, in the coupling scheme the single unit energy and its variance are plotted as a function of $\tilde{g}\tau\sqrt{N}$, while they are plotted as a function of $\tilde{g}\tau$ in the detuning case.}
        \label{fig:energy_inv_time}
    \end{minipage}
\end{figure}

In this appendix we further comment on the scaling of the charging time that we discussed in the main text, i.e. that we observe no collective scaling in the coupling scheme, while we do observe it in the detuning charging scheme. 
To further substantiate this claim, in Figs.~\ref{fig:ergo_inv_time} and \ref{fig:energy_inv_time} we respectively show the same exact plots as in Figs.~1 and 2 of the main text, the only difference being the scaling of the time in the x-axis, which is inverted between the charging and detuning schemes. By direct comparison, we see that such an inversion produces charging curves that do no overlap, while they do with the scaling reported in the main text. This suggest that the scaling reported in the main text is correct.

We now comment on the choice of the time-step $\Delta t$ reported in the caption of Fig.~1 of the main text. As commented in the main text, we chose a fixed $\Delta t$ in the coupling scheme, and one scaling as $\sim 1/\sqrt{N}$ in the detuning scheme, to follow the scaling of the charging time in the respective cases. For completeness, we tried optimizing the coupling scheme using the same choice of $\Delta t$ that we used for the detuning case. This actually yielded the possibility of reaching higher values of the ergotropy than the ones reported in this manuscript, but at the expense of a high injection of energy through the driving. However, no clear scaling of the charging time was visible. Therefore, in the present manuscript we decided to report the results for fixed $\Delta t$ which, notably, yield a nearly fully-charged battery without any injection of external energy through the driving.
Interestingly, this is not a numerical optimization error (whose robustness is discussed in Sec.~\ref{app:rl_robust}), rather it has a physical origin. By decreasing $\Delta t$, we allow faster driving schemes, i.e. higher frequencies in the driving control. This allows the RL agent to increase the ergotropy by exploiting a fast modulation of the control that injects energy into the system. Indeed, in the detuning case, which has a smaller $\Delta t$, we find a collective speedup of the charging time at the expense of injecting a large amount of energy. Conversely, in the coupling scheme we do not find a collective speedup, but we find charging protocol with nearly no energy injected from the driving.

\section{Non-greedy RL charging protocols}
\label{app:protocols}
\begin{figure}[!tb]
	\centering	\includegraphics[width=0.55\columnwidth]{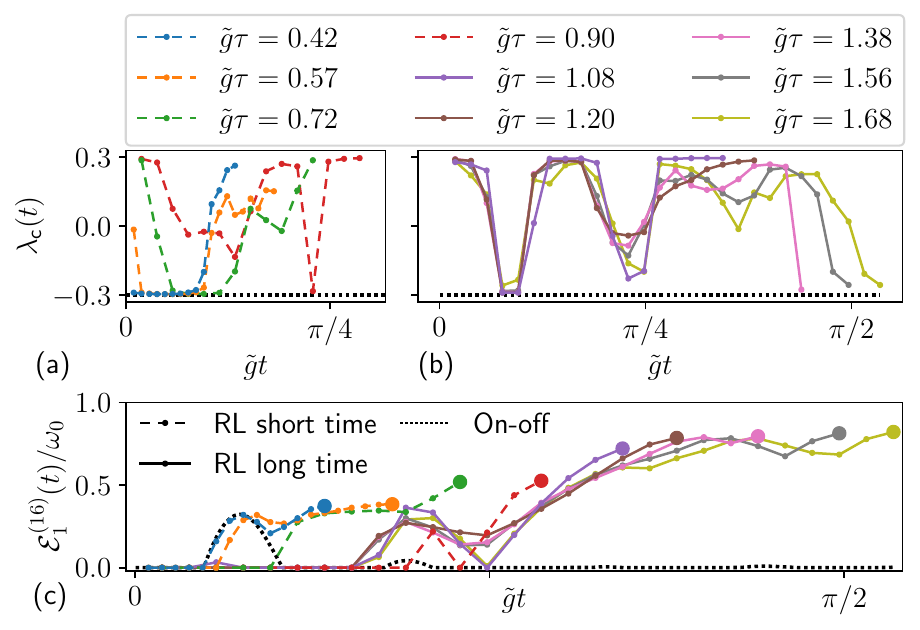}
	\caption{Optimal charging protocol for the coupling scheme. $\lambda_\text{c}(t)$ is plotted, as a function of time $\tilde{g}t$, for short (a) and long (b) charging times that produce the results of Figs.~\ref{fig:sketch} and \ref{fig:energy} for $N=16$ TLSs. Each colored line corresponds to a distinct RL optimization with different charging time $\tau$. The corresponding ergotropy of a single battery unit $\mathcal{E}^{(16)}_1(t)/\omega_0$ is shown in (c). The small dots in (a,b) correspond to values of the control determined by RL at each time-step, while the large dots in (c) correspond to the ergotropy at the final time $\tau$; these are the values reported in Figs.~\ref{fig:sketch} and \ref{fig:energy}. The black-dotted lines corresponds to on-off protocols.}
	\label{fig:coupling_control}
\end{figure}
\begin{figure}[!tb]
	\centering	\includegraphics[width=0.55\columnwidth]{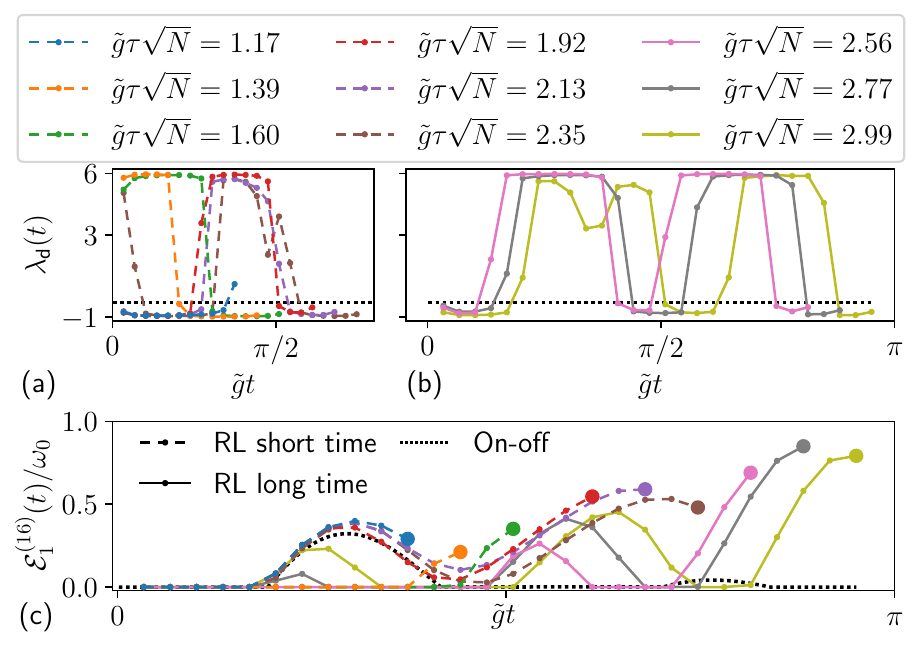}
	\caption{Optimal charging protocol for the detuning scheme plotted as in Fig.~\ref{fig:coupling_control} for the coupling scheme.}
	\label{fig:detuning_control}
\end{figure}

In this appendix we show and discuss the charging protocols that emerge from the RL optimization, and we explicitly show how these come from a non-greedy optimization.

In Figs.~\ref{fig:coupling_control} and \ref{fig:detuning_control} we analyze respectively the charging protocols $\lambda_\text{c}(t)$ and $\lambda_\text{d}(t)$ discovered by the RL method that produce the results shown in Figs.~\ref{fig:sketch} and \ref{fig:energy}. We plot a different curve for each value of the charging time $\tau$ (each one corresponds to a separate RL optimization), and we consider $N=16$ (similar findings hold for other values of $N$).  
For clarity, we separately display the charging protocols for short [panel (a), dashed lines] and long [panel (b), full lines] charging time $\tau$. 
Panel (c) reports the corresponding ergotropy $\mathcal{E}_1^{(16)}(t)$ for each protocol shown in panels (a,b). The thick dots at the end of the curves represent the values of $\mathcal{E}_1^{(16)}(\tau)$ delivered at the final time $\tau$; these correspond to the values shown in Fig.~\ref{fig:sketch}(b,c) along the $N=16$ curve. The black-dotted lines in Figs.~\ref{fig:coupling_control} and \ref{fig:detuning_control} correspond to the on-off protocol.

While some general trends can be seen, e.g. the curves in Figs.~\ref{fig:coupling_control}(a) and \ref{fig:detuning_control}(a) share some features respectively with Figs.~\ref{fig:coupling_control}(b) and \ref{fig:detuning_control}(b), in general they do not overlap.
This signals that the optimal charging strategy is \textit{non-greedy}. While a greedy strategy chooses values of the control that, at every time $t$, maximize the instantaneous increase of the ergotropy, in Fig.~\ref{fig:coupling_control}(c) and Fig.~\ref{fig:detuning_control}(c) we see that the charging curves for large values of $\tau$ have a lower ergotropy for short times than protocols with smaller $\tau$. This short-term sacrifice is what allows them to reach a larger ergotropy at the final time $\tau$. Equivalently, the non-greediness is signalled by the fact that the final ergotropy, shown as thick circles, lies substantially above the other curves obtained for larger values of $\tau$.
This shows that optimal charging strategies are non-trivial and generally depend on the charging time $\tau$.

\section{Details on the Reinforcement Learning Method}
In this appendix we first provide in Sec.~\ref{app:rl_setting} a general explanation of what Reinforcement Learning (RL) is (for an in-depth explanation of RL, we refer to Ref.~\cite{sutton2018}). 
We then explain in Sec.~\ref{app:rl_batteries} how we apply this method to the optimal charging of quantum batteries, and in Sec.~\ref{app:sac} we provide details on the specific algorithm we used, namely the soft actor-critic method \cite{haarnoja2018,haarnoja2019}.
In Sec.~\ref{app:rl_implementation} we provide implementation details, such as the neural network architecture, the training method, and the value of the hyperparameters, used to find the results presented in the main text, and in Sec.~\ref{app:rl_robust} we discuss the robustness of the method.

\subsection{Reinforcement Learning Setting}
\label{app:rl_setting}
\setcounter{equation}{18}%
Reinforcement Learning is a general tool, based on the Markov decision process framework \cite{sutton2018}, that can tackle optimization problems formulated in the following way. A \textit{computer agent} must learn to master some task by repeatedly interacting with an \textit{environment}. 
Let us consider the time interval $[0,\tau]$ and discretize time in time-steps of duration $\Delta t = \tau/(M-1)$, such that the discrete times $t_i = i\Delta t$ span the time interval $[0,\tau]$ for $i\in\{0,1,\dots, M-1\}$.
Let us denote with $s_i\in \mathcal{S}$ the state of the environment at time $t_i$, where $\mathcal{S}$ is the \textit{state space}.
At every time-step, the agent chooses an action $a_i\in\mathcal{A}$ to perform on the environment, where $\mathcal{A}$ is the \textit{action space}. The action is chosen by sampling it from the \textit{policy function} $\pi(a_i|s_i)$, which describes the probability density of choosing action $a_i$, provided that the environment is in state $s_i$. The environment reacts to the chosen action by returning to the computer agent the new state $s_{i+1}$ at the following time-step, and returning a \textit{reward} $r_{i+1}$ which is a scalar quantity. The Markov decision process assumption requires that the the state $s_{i+1}$ and the reward $r_{i+1}$ must only depend (eventually stochastically) on the last state $s_i$ and on the last chosen action $a_i$.

In this manuscript, we consider the \textit{episodic setting}. An ``episode'' starts at $t_0 = 0$ in a reference state $s_0=\sigma_0$, and ends at $t_{M-1}=\tau$ after $M$ steps. the goal of RL is then to learn an optimal policy $\pi^*(a|s)$ that maximizes the expected \textit{return} $g_0$ i.e. the sum of the rewards
\begin{equation}
    g_0 = r_{1} + \gamma r_{2} + \gamma^2 r_{3} + \dots + \gamma^{M-2}r_{M-1}  =  \sum_{k=0}^{M-2} \gamma^{k} r_{k+1},
    \label{eq:g0_def}
\end{equation}
where $\gamma\in[0,1]$ is the so-called ``discount factor'' which determines how much we privilege short or long-term rewards.
An optimal policy is thus defined as
\begin{equation}
    \pi^* = \mathrm{arg}\max_\pi\, \mathrm{E}_\pi  \Big[ g_0 \Big| s_0 = \sigma_0 \Big],
    \label{eq:pi_star_rl}
\end{equation}
where the expectation value $E_\pi[\cdot]$ in Eq.~(\ref{eq:pi_star_rl}) is taken with respect to the stochasticity in the choice of the actions according to the policy $\pi$, and with respect to the state evolution of the environment.

Starting from a random policy, and repeating many episodes over and over, the RL algorithm should learn an optimal policy. How learning takes place depends on the specific RL algorithm. As detailed below, in this manuscript we use the soft-actor critic method, proposed in Refs.~\cite{haarnoja2018, haarnoja2019}, with a few modifications that will be detailed throughout this appendix. We thus refer to Refs.~\cite{haarnoja2018, haarnoja2019} for further details of the method.

\subsection{RL for quantum batteries}
\label{app:rl_batteries}
We now detail how we apply the RL framework to optimize the final ergotropy $\mathcal{E}^{(N)}_1(\tau)$. 
As state $s_i$ of the environment, we choose the wave-function $\ket{\psi(t_i)}$ of the charger and battery system combined at time $t_i$, together with the last chosen action, and the current time-step. In particular, we expand the wave-function in the product basis of Fock states for the photonic mode (truncated up to a maximum number of photons $N_\text{Fock}$), and of the Dicke basis for the two-level systems defined in App.~\ref{app:ergotropy}. We then take the real and imaginary part of each coefficient, stack them into a vector, append the last action and the current time-step, and use this as state. The initial state $\sigma_0$ encodes the state $\ket{\psi(0)}=\ket{{\rm G}}\otimes\ket{N}$ defined in the main text.

As action $a_i$, we choose the value of the control $\lambda_\text{c}(t)$ or $\lambda_\text{d}(t)$ that will then be kept constant in the time interval $[t_i, t_{i+1}]$. This will end up constructing a piece-wise constant charging protocol. The action can be any value in a continuous interval.

As reward $r_{i+1}$, we choose the variation in ergotropy
\begin{equation}
    r_{i+1} = \mathcal{E}^{(N)}_1(t_{i+1}) - \mathcal{E}^{(N)}_1(t_{i}),
\end{equation}
such that the return $g_0$, which is the quantity being optimized by RL, is given by 
\begin{equation}
    g_0 = r_1 + \dots + r_{M-1} = \mathcal{E}^{(N)}_1(\tau),
\end{equation}
provided that we choose $\lambda=1$. Notice that $\mathcal{E}^{(N)}_1(t_0) = 0$ since we start from a totally discharged state.

This choice of state and reward respects the Markov decision process assumption. Indeed, using the Schr{\"o}dinger equation, we can compute the state $s_{i+1}$ simply knowing $s_i$ and $a_i$, and the reward is also just a function of $s_i$ and $a_i$ since it can be computed from $s_i$ and $s_{i+1}$.

\subsection{Soft actor-critic algorithm}
\label{app:sac}
The soft actor-critic (SAC) algorithm \cite{haarnoja2018,haarnoja2019} starts from a random policy and iteratively improves it until an optimal (or near-optimal) policy is reached. The method is based on policy iteration, i.e. it consists of iterating over two steps: a \textit{policy evaluation step}, and a \textit{policy improvement step.}
In the policy evaluation step, the quality of the current policy is evaluated by estimating the \textit{value function} $Q^\pi(s,a)$, while in the policy improvement step a better policy is found making use of the value function. Before elaborating on these two steps, we introduce some notion that will be used later on, and we provide a definition of the value function $Q^\pi(s,a)$.

In the SAC method, balance between exploration and exploitation \cite{sutton2018} is achieved by introducing an entropy-regularized maximization objective. Instead of defining an optimal policy according to Eqs.~(\ref{eq:g0_def}) and (\ref{eq:pi_star_rl}), an optimal policy is defined as
\begin{equation}
    \pi^* = \mathrm{arg}\max_\pi\, \mathrm{E}_\pi\Big[
    \sum_{k=0}^{M-2} \gamma^k \,\Big(r_{k+1} + \alpha H[\pi(\cdot|s_k)]  \Big) \Big| s_0 =\sigma_0  \Big],
    \label{eq:pi_star_0}
\end{equation}
where $\alpha \geq 0$ is known as the ``temperature'' parameter that balances the trade-off between exploration and exploitation, and
\begin{equation}
    H[P] = \mathop{\mathrm{E}}\limits_{x\sim P}[ -\log P(x) ]
\end{equation}
is the entropy of the probability density $P(x)$. Notice that Eq.~(\ref{eq:pi_star_0}), for $\alpha=0$, reduces to the previous definition of optimal policy given in Eq.~(\ref{eq:pi_star_rl}). A positive value of $\alpha$ will favour a more exploratory behaviour, since a higher entropy distribution is less deterministic.
For notation simplicity, we now assume that information about the current time $t$ is encoded in the state $s$. We then adopt the convention that both $r_{k+1}=0$ and $H[\pi(\cdot|s_k)]=0$ if the state $s_k$ has reached time $t=\tau$. Furthermore, using the Markov decision process assumption, we notice that an optimal policy also maximizes the sum of the future rewards starting from any intermediate states - not only from the initial state. Therefore, we write an optimal policy as
\begin{equation}
    \pi^* = \mathrm{arg}\max_\pi\, \mathop{\mathrm{E}_\pi}\limits_{s\sim \mu_\pi}\Big[ \sum_{k=0}^{\infty} \gamma^k \,\Big(r_{k+1} + \alpha H[\pi(\cdot|s_k)]  \Big) \Big| s_0 =s  \Big].
    \label{eq:pi_star_1}
\end{equation}
As opposed to Eq.~(\ref{eq:pi_star_0}), we now extend the sum to infinity (thanks to the encoding of time into the state and the conventions introduced above), and we sample the initial state $s$ from the steady-state distribution of states $\mu_\pi$ that are visited starting from the initial state $s_0=\sigma_0$, and then choosing actions according to the policy $\pi$. At last, since the distribution $\mu_\pi$ would be difficult to calculate in practice, we replace it with $\mathcal{B}$, which is a replay buffer populated during training by storing the observed one-step transitions $(s_k,a_k, r_{k+1}, s_{k+1})$. We thus arrive to
\begin{equation}
    \pi^* = \mathrm{arg}\max_\pi\, \mathop{\mathrm{E}_\pi}\limits_{s\sim \mathcal{B}}\Big[ \sum_{k=0}^{\infty} \gamma^k \,\Big(r_{k+1} + \alpha H[\pi(\cdot|s_k)]  \Big) \Big| s_0 =s  \Big].
    \label{eq:pi_star}
\end{equation}

Equation~(\ref{eq:pi_star}) is now our optimization objective. Accordingly, we define the value function as
\begin{equation}
    Q^\pi(s,a) = 
    \text{E}_\pi  \left[ r_{1} + 
    \sum_{k=1}^{\infty} \gamma^k \,\Big(r_{k+1} + \alpha H[\pi(\cdot|s_k)]  \Big) \Big| s_0=s, a_0=a \right].
    \label{eq:q_def}
\end{equation}
Its recursive Bellman equation therefore reads
\begin{equation}
    Q^\pi(s,a) =  \underset{s_1 \atop a_1 \sim \pi(\cdot|s_1) }{\text{E}} \Big[ r_{1} + 
    \gamma \Big( Q^\pi(s_1,a_1) + \alpha H[\pi(\cdot|s_1)]   \Big) \Big| s_0=s, a_0=a \Big].
    \label{eq:bellman}
\end{equation}
$Q^\pi(s,a)$ is thus the weighed sum of future rewards that one would obtain starting from state $s$, performing action $a$, and choosing all subsequent actions according to the policy $\pi$. It plays the role of a ``critic'' that judges the quality of the actions chosen according to the policy $\pi$, which plays the role of an ``actor''.

We now focus on the policy. Here, we assume the action to be a single continuous action lying in the interval $[a_1,a_2]$, although a generalization to multiple continuous actions is straightforward. As in Refs.~\cite{haarnoja2018,haarnoja2019}, we parameterize $\pi(a|s)$ as a squashed Gaussian policy, i.e. as the distribution of the variable
\begin{align}
 \tilde{a}(\xi|s) &= 
    a_1 + \frac{a_2 - a_1}{2}[1+ \tanh\left( \mu(s) + \sigma(s)\cdot \xi )  \right)],   &  \xi &\sim \mathcal{N}(0,1),
    \label{eq:a_tilda}
\end{align}
where $\mu(s)$ and $\sigma(s)$ represent respectively the mean and standard deviation of the Gaussian distribution, and $\mathcal{N}(0,1)$ is the normal distribution with zero mean and unit variance. This is the so-called reparameterization trick.

We now describe the policy evaluation step. In the SAC algorithm, we learn two value functions $Q_{\phi_i}(s,a)$ described by a set of learnable parameters $\phi_i$, for $i=1,2$. $Q_\phi(s,a)$ is a function approximator, e.g. a neural network, that will be determined minimizing a loss function. 
Since $Q_{\phi_i}(s,a)$ should satisfy the Bellman Eq.~(\ref{eq:bellman}), we define the loss function for $Q_{\phi_i}(s,a)$ as the mean square difference between the left and right hand side of Eq.~(\ref{eq:bellman}), i.e.
\begin{equation}
    L_Q(\phi_i) = \mathop{\mathrm{E}}\limits_{(s,a,r,s^\prime)\sim \mathcal{B}} \left[ ( Q_{\phi_i}(s,a) - y(r,s^\prime))^2  \right],
    \label{eq:q_loss}
\end{equation}
where 
\begin{equation}
    y(r,s^\prime) = r +  \gamma \underset{a^\prime \sim \pi(\cdot|s^\prime)}{\text{E}} \Big[ \min_{j=1,2}Q_{\phi_{\text{targ},j}}(s^\prime,a^\prime) + \alpha H[\pi(\cdot|s^\prime)] \Big].
    \label{eq:y_1}
\end{equation}    
Notice that in Eq.~(\ref{eq:y_1}) we replaced $Q^\pi$ with $\min_{j=1,2}Q_{\phi_{\mathrm{targ},j}}$, where $\phi_{\mathrm{targ},j}$, for $j=1,2$, are target parameters which are not updated when minimizing the loss function; instead, they are held fixed during backpropagation, and then they are updated according to Polyak averaging, i.e.
\begin{equation}
    \phi_{\mathrm{targ},i} \leftarrow \rho_\mathrm{polyak} \phi_{\mathrm{targ},i} + (1-\rho_\mathrm{polyak})\phi_{i},
\end{equation}
where $\rho_\mathrm{polyak}$ is a hyperparameter. This change was shown to improve learning \cite{haarnoja2018,haarnoja2019}. Writing the entropy explicitly as an expectation values, we have
\begin{equation}
    y(r,s^\prime) = r + \gamma   
      \mathop{\mathrm{E}}\limits_{a^\prime \sim \pi(\cdot|s^\prime)}  \left[  \min_{j=1,2}Q_{\phi_{\text{targ},j}}(s^\prime,a^\prime) - \alpha\log\pi(a^\prime|s^\prime) \right].
    \label{eq:y_2}
\end{equation}
We then replace the expectation value over $a^\prime$ in Eq.~(\ref{eq:y_2}) with a single sampling $a^\prime\sim\pi(\cdot|s^\prime)$ performed using Eq.~(\ref{eq:a_tilda}). 

We now turn to the policy improvement step. Let $\pi_\theta(a|s)$ be a parameterization of the policy function that depends on a set of learnable parameters $\theta$. In particular, the functions $\mu_\theta(s)$ and $\sigma_\theta(s)$ defined in Eq.~(\ref{eq:a_tilda}) will be parameterized using neural networks. Given a policy $\pi_{\theta_{\mathrm{old}}}(a|s)$, Refs.~\cite{haarnoja2018,haarnoja2019} prove that $\pi_{\theta_{\mathrm{new}}}(a|s)$ is a better policy [with respect to maximization in Eq.~(\ref{eq:pi_star})] if we update the policy parameters according to
\begin{equation}
    \theta_\mathrm{new} = \mathrm{arg}\min_\theta D_\mathrm{KL}\Big( \pi_\theta(\cdot|s)   \Big|\Big| \frac{\exp\left( Q^{\pi_{\theta_\mathrm{old}}}(s,\cdot)/\alpha \right) }{ Z^{\pi_{\theta_\mathrm{old}}}}  \Big),
    \label{eq:theta_new}
\end{equation}
where $s$ is any state, $D_\mathrm{KL}$ denotes the Kullback-Leibler divergence, and $Z^{\pi_{\theta_\mathrm{old}}}$ is the partition function of the exponential of the value function. Conceptually, this step is similar to making the policy $\epsilon$-greedy in the standard RL setting. The idea is to use the minimization in Eq.~(\ref{eq:theta_new}) to define a loss function to perform an update of $\theta$. Noting that the partition function does not impact the gradient, multiplying the Kullback-Leibler divergence by $\alpha$, and replacing $Q^{\pi_{\theta_\mathrm{old}}}$ with $\min_j Q_{\phi_j}$, we define the loss function as
\begin{equation}
    L_\pi(\theta) = \mathop{\mathrm{E}}\limits_{{s\sim \mathcal{B} \atop   a\sim \pi_\theta(\cdot|s)}}\left[ \alpha \log\pi_\theta(a|s) - \min_{j=1,2} Q_{\phi_j}(s,a) \right].
    \label{eq:pi_loss}
\end{equation}
As before, in order to evaluate the expectation value in Eq.~(\ref{eq:pi_loss}), we replace the expectation value over $a$ with a single sampling $a^\prime\sim\pi(\cdot|s^\prime)$ performed using Eq.~(\ref{eq:a_tilda}). 

We have defined and shown how to evaluate the loss functions $L_Q(\phi)$ and $L_\pi(\theta)$ that allow us to determine the value function and the policy [see Eqs.~(\ref{eq:q_loss}), (\ref{eq:y_2}) and (\ref{eq:pi_loss})]. Now, we discuss how to automatically tune the temperature hyperparameter $\alpha$. Ref.~\cite{haarnoja2019} shows that constraining the average entropy of the policy to a certain value leads to the same exact same SAC algorithm, with the addition of an update rule to determine the temperature. Let $\bar{H}$ be the fixed average values of the entropy of the policy. We can then determine the temperature $\alpha$ minimizing the following loss function
\begin{equation}
    L_\text{temp}(\alpha) = \alpha \underset{ \substack{s\sim \mathcal{B} }}{\text{E}}\left[ H[\pi(\cdot|s)] - \bar{H} \right] = \alpha \underset{ \substack{s\sim \mathcal{B} \\ a^\prime \sim \pi(\cdot|s) }}{\text{E}}\left[ -\ln\pi(a^\prime|s) - \bar{H} \right].
\label{eq:l_temp}
\end{equation}
As usual, we replace the expectation value over $a^\prime$ with a single sampling $a^\prime\sim\pi(\cdot|s^\prime)$ performed using Eq.~(\ref{eq:a_tilda}). 

To summarize, the SAC algorithm consists of repeating over and over a policy evaluation step, a  policy improvement step, and a step where the temperature is updated. The policy evaluation step consists of a single optimization step to minimize the loss functions $L_Q(\phi_i)$ (for $i=1,2$), given in Eq.~(\ref{eq:q_loss}), where $y(r,s^\prime)$ is computed using Eq.~(\ref{eq:y_2}).
The policy improvement step consists of a single optimization step to minimize the loss function $L_\pi(\theta)$ given in Eq.~(\ref{eq:pi_loss}). The temperature is then updated performing a single optimization step to minimize $L_\text{temp}(\alpha)$ given in Eq.~(\ref{eq:l_temp}). In all loss functions, the expectation values with respect to $\mathcal{B}$ are approximated with a batch of experience sampled randomly from the replay buffer $\mathcal{B}$, and the expectation values with respect to the action $a^\prime$ are replaced with a single sampling $a^\prime\sim\pi(\cdot|s^\prime)$ performed using Eq.~(\ref{eq:a_tilda}).

\subsection{RL implementation details and training hyperparameters}
\label{app:rl_implementation}
Here we provide details about the RL implementation and the hyperparameters used for training. Notice that, in all trainings, regardless of the number of qubits $N$, we use nearly the same hyperparameters.

Both the policy function and the value function are parameterized using fully-connected neural networks with $2$ hidden layers, and using the ReLU activation function in all layers except for the output layer that is linear. We further normalize the input to both neural networks such that it lies in the interval $[-\sqrt{12},\sqrt{12}]$. This guarantees that, if the input was uniformly distributed in such interval, it would have unit variance.

The value function $Q(s,a)$ takes as input the state $s$ and the action $a$ stacked together. They are normalized assuming that the real and imaginary parts of the coefficients of the the wave-function expansion lie in $[-1,1]$, that time lies in $[0,\tau]$, and that the last action lies in the interval $[a_1, a_2]$. The neural network then outputs a single value representing the value function $Q(s,a)$.

The policy function $\pi(a|s)$ is parameterized by a neural network that takes the state $s$ as input (normalized as for the value function), and outputs two values, $\mu(s)$ and $m(s)$. $\mu(s)$ represents the mean of the Gaussian, defined in Eq.~(\ref{eq:a_tilda}), while the variance is computed as $\sigma(s) = m^2 + 10^{-7}$. This guarantees that the variance will be non-negative.

Training occurs by repeating many episodes, each of which is made up of $M$ time-steps. We denote with $n_\text{steps}$ the total number of time-steps performed during the whole training, thus across all episodes. As in Ref.~\cite{Erdman1}, to enforce sufficient exploration in the early stage of training, we do the following. For a fixed number of initial steps $n_\text{init-rand}$, we choose random actions sampling them uniformly withing their range. Furthermore, for another fixed number of initial steps $n_\text{init-no-upd}$, we do not update the neural network parameters to allow the replay buffer to have enough transitions. $\mathcal{B}$ is a first-in-first-out buffer, of fixed dimension, that is populated with the observed transitions $(s_k,a_k,r_{k+1},s_{k+1})$.  Batches of transitions are then randomly sampled from $\mathcal{B}$ to compute the loss functions and update the neural network parameters. After this initial phase, we repeat a policy evaluation, a policy improvement step  and a temperature update step $n_\text{updates}$ times every $n_\text{updates}$ steps (a step being a choice of the action according to the policy function, or randomly in the initial training phase). This way, the overall number of updates coincides with the total number of actions performed (across all episodes). The optimization steps for the value function and the policy are performed using the ADAM optimizer with the standard values of $\beta_1$ and $\beta_2$, and learning rate $\text{LR}$. The temperature parameter $\alpha$ is determined using stochastic gradient descent with learning rate $\text{LR}_\alpha$. To favor an exploratory behavior early in the training, and at the same time to end up with a policy that is approximately deterministic, we schedule the target entropy $\bar{H}$. In particular, we vary it exponentially at each time-step during training as
\begin{equation}
    \bar{H}(n_\text{steps}) = \bar{H}_{\text{end}} 
    + (\bar{H}_{\text{start}}-\bar{H}_{\text{end}})\exp(-n_\text{steps}/\bar{H}_{\text{decay}}),
\end{equation}
where $\bar{H}_{\text{start}}$, $\bar{H}_{\text{end}}$ and $\bar{H}_{\text{decay}}$ are hyperparameters. Furthermore, in order to have hyperparameters that are less environment-dependent, instead of computing the entropy $H[\pi(\cdot|s)]$ of the policy, we compute the entropy of the policy as if it outputted values in a fixed reference interval $[-1,1]$. In practice, this is implemented computing $\ln\pi(a|s)$ in all loss functions making this assumption. It can be seen that this variation simply amounts to an additive constant.

\begin{table}[htb]
\centering
\begin{tabular}{lccc}
\toprule
Hyperparameter ~ & ~ Value (coupling scheme) ~ & ~ Value (detuning scheme)  \\
\midrule
Batch size  & 256 & ''  \\
Training steps & 480k & '' \\
$\text{LR}$ & 0.001 & ''  \\
$\text{LR}_\alpha$ & 0.003 & ''   \\
$\gamma$ & 0.993 & '' \\
$\mathcal{B}$ size  & 180k & ''   \\
$\rho_\text{polyak}$ & 0.995 & ''  \\
Units in first hidden layer & 512 & '' \\
Units in second hidden layer & 256 & '' \\
$n_\text{init-rand}$  & 5k & '' \\
$n_\text{init-no-update}$  & 1k & '' \\
$n_\text{updates}$  & 50 & ''   \\
$\bar{H}_{\text{start}}$ & 0.72 & ''  \\
$\bar{H}_{\text{end}}$ & -3.0 & ''  \\
$\bar{H}_{\text{decay}}$ & 200k & ''  \\
$c_\text{mean}$ & 40k & 60k  \\
$c_\text{width}$ & 20k & ''  \\
$N_\text{Fock}$ & $2N$ & $5N$ \\
\bottomrule
\end{tabular}
\caption{Hyperparameters used in all numerical calculations reported in this manuscript. Letter ``k'' stands for thousand, and the quotes symbol in the detuning scheme column means the same values as in the coupling scheme.}
\label{tab:hyper}
\end{table}
To enforce that the temperature parameter $\alpha$ never accidentally becomes negative during training, instead of minimizing directly $L_\text{temp}(\alpha)$ given in Eq.~(\ref{eq:l_temp}), we parameterize the temperature in terms of a parameter $l_\alpha$ as $\alpha(l_\alpha)=e^{l_\alpha}$, and we determine $l_\alpha$ minimizing the loss function $L_\text{temp}(\alpha(l_\alpha))$.

At last, we use an additional trick during the initial part of the training to start learning a meaningful policy. As can be seen in the main text, even under optimal control, the ergotropy remain exactly zero for a considerable amount of time. This means that, especially during the early phases of training when the policy is still random, the RL agent is constantly receiving zero reward. In order to initially drive the agent towards a better policy, we first use the energy difference of the battery as reward, and then we smoothly change it back to the ergotropy difference during training. More specifically, we use as reward
\begin{equation}
    r_{i+1} = c(n_\text{steps}) \frac{ E_1^{(N)}(t_{i+1}) - E_1^{(N)}(t_{i})}{\omega_0} + (1-c(n_\text{steps}))\frac{  \mathcal{E}^{(N)}_1(t_{i+1}) - \mathcal{E}^{(N)}_1(t_{i})}{\omega_0},
\end{equation}
where
\begin{equation}
    c(n_\text{steps}) = \left(1 + e^{ (n_\text{steps}-c_\text{mean})/c_\text{width} }\right)^{-1},
\end{equation}
and where $c_\text{mean}$ and $c_\text{width}$ are hyperparameters. Essentially, during training we switch from optimizing the energy to optimizing the ergotropy using a weight proportional to the Fermi distribution centered around $c_\text{mean}$ with characteristic width $c_\text{width}$.

All hyperparameters used to produce the results in this manuscript are provided in Table~\ref{tab:hyper}.
The only difference between the coupling and the detuning scheme is in $c_\text{mean}$ and in $N_\text{Fock}$. The larger Fock space was used in the detuning scheme since more energy is injected into the system [see Fig.~1(d,e) of the main text], and the larger $c_\text{mean}$ gave a slightly better convergence.

We verified that convergence in the cutoff size $N_\text{Fock}$ of the Fock space is reached, and we report all quantities using $N_\text{Fock}=6N$ and $N_\text{Fock}=10N$ during evaluation respectively in the coupling and detuning cases.

We conclude commenting the wall time necessary to run the RL method. We ran our simulations on a desktop computer using an NVIDIA GeForce RTX 3090 as GPU. Higher values of $N$ are slower to train because they use larger neural networks. For $N=12$, we could run $5$ optimizations at the same time, requiring $45$ minutes per optimization. For $N=20$, we could only run $3$ optimizations at the same time (due to memory limitations), requiring $73$ minutes per optimization.

\subsection{Robustness of the RL results}
\label{app:rl_robust}
\begin{figure}[!htb]
	\centering
	\includegraphics[width=0.55\columnwidth]{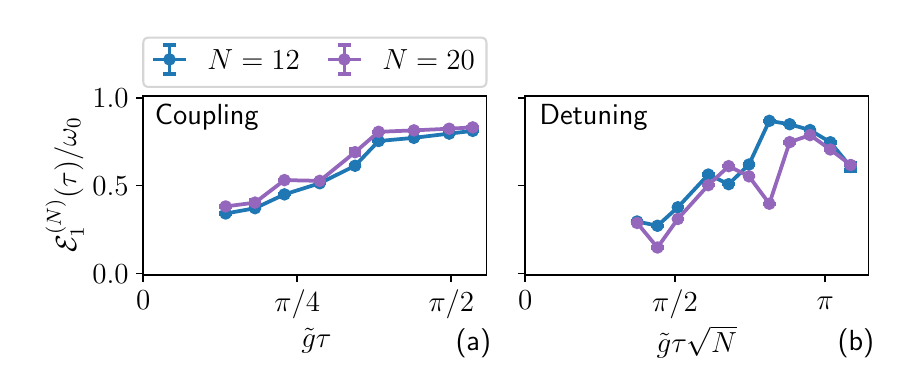}
	\caption{Average (as dots) and standard deviation (as error bars) of the final ergotropy, computed over $5$ repetitions of the RL optimization, as a function of the charging time $\tilde{g}\tau$ in the coupling scheme [panel (a)] and of the rescaled charging time $\tilde{g}\tau\sqrt{N}$ in the detuning scheme [panel (b)]. The best of the five optimizations is reported in the main text. Only $N=12$ and $N=20$ are reported here to make the dots and corresponding error bars more visible. The system parameters and plotting style are the same as in Fig.~1(b,c) of the main text.
 }
\label{fig:convergence}
\end{figure}
In this subsection we discuss the robustness of the optimization method. All optimizations carried out were repeated $5$ times, and the repetition with the largest final ergotropy is shown in the Figures of the main text. 
Notably, every repetition of the optimization provided results that are very similar to one another. 
Indeed. in Fig.~\ref{fig:convergence} we plot the average (as dots) and the standard deviation (as error bars) of the ergotropy over the $5$ repetitions. This is plotted in the same style and scale as Figs.~1(b,c) of the main text, both in the coupling and detuning scheme. As we can see, the error bars are hardly visible on this scale, except for one dot along the $N=20$ curve in the coupling scheme, and for the final time point in the detuning case; in either case the error bars are barely visible. This demonstrates the stability of the RL optimization method (only $N=12$ and $N=20$ are shown in Fig.~\ref{fig:convergence} to make the dots and corresponding error bars more visible. However, the same holds also for the intermediate values of $N$).

\end{document}